\documentclass[prl,twocolumn,superscriptaddress]{revtex4-1}
\usepackage{amsmath}
\usepackage{amssymb}
\usepackage{graphics}
\usepackage{graphicx}
\usepackage{float}
\usepackage{subfigure}
\usepackage{color}
\usepackage{hyperref}
\begin{document}
\title{Bulk and edge dynamics of a 2D Affleck-Kennedy-Lieb-Tasaki model}

\author{Zenan Liu}
\affiliation{State Key Laboratory of Optoelectronic Materials and Technologies, Center for Neutron Science and Technology, School of Physics, Sun Yat-sen University,
Guangzhou, 510275, China}

\author{Jun Li}
\affiliation{State Key Laboratory of Optoelectronic Materials and Technologies, Center for Neutron Science and Technology, School of Physics, Sun Yat-sen University,
Guangzhou, 510275, China}
\affiliation{Key Laboratory for Microstructural Material Physics of Hebei Province, School of Science, Yanshan University, Qinhuangdao 066004, China}

\author{Rui-Zhen Huang}
\affiliation{Kavli Institute for Theoretical Sciences, University of Chinese Academy of Sciences, Beijing 100190, China}

\author{Jun Li}
\affiliation{State Key Laboratory of Optoelectronic Materials and Technologies, Center for Neutron Science and Technology, School of Physics, Sun Yat-sen University,
Guangzhou, 510275, China}

\author{Zheng Yan}
\email{zhengyan@hku.hk}
\affiliation{Beihang Hangzhou Innovation Institute Yuhang, Hangzhou 310023, China}
\affiliation{Department of Physics and HKU-UCAS Joint Institute of Theoretical and Computational Physics,
The University of Hong Kong, Pokfulam Road, Hong Kong, China}
\affiliation{State Key Laboratory of Surface Physics and Department of Physics, Fudan University, Shanghai 200438, China}

\author{Dao-Xin Yao}
\email{yaodaox@mail.sysu.edu.cn}
\affiliation{State Key Laboratory of Optoelectronic Materials and Technologies, Center for Neutron Science and Technology, School of Physics, Sun Yat-sen University,
Guangzhou, 510275, China}

\begin{abstract}
  We study the dynamical properties of both bulk and edge spins of a two-dimensional Affleck-Kennedy-Lieb-Tasaki (AKLT) model mainly by using the stochastic series expansion quantum Monte Carlo method with stochastic analytic continuation. In the deep AKLT phase, we obtain a spin spectrum with the flat band, which is a strong evidence for a localized state. Through the spectrum analysis, we see a clear continuous phase transition from the AKLT phase to the N\'eel phase in the model, and the energy gap becomes closed at the corresponding momentum point. In comparison with linear spin-wave theory, the differences show that there are strong interactions among magnons at high energies. With an open boundary condition, the gap of edge spins in the AKLT phase closes at both the $\Gamma$ point and the $\pi$ point interestingly to emerge into a flat-band-like Luttinger liquid phase, which can be explained by symmetry and perturbation approximation. This paper helps us to better understand the completely different dynamical behaviors of bulk and edge spins in the symmetry protected topological phase.
\end{abstract}
\date{\today}
\maketitle

%%%%%%%%%%%%%%%%%%%%%%%%%%%%%%%%%%%%%%%%%%%%%%%%%%%%%%%%%%%%

\section{Introduction}

The Haldane phase \cite{1983Continuum,1983Nonlinear} is a famous symmetry protected topological (SPT) phase which only has short-range quantum entanglement and is first found in the $S=1$ Heisenberg antiferromagnetic chain. Generally, the SPT phase is proposed to describe the system protected by symmetry in any dimension that can not be transformed into a trivial state unless closing the bulk gap or breaking the symmetry\cite{2009Tensor}.  Recently, some numerical results have found that the non trivial surface in the SPT phase can cause unusual critical behaviors, whose mechanism is still puzzling and under debate \cite{zhang2017unconventional,ding2018engineering,zhu2021surface,weber2018nonordinary}. In a previous work, the two-dimensional Affleck-Kennedy-Lieb-Tasaki (AKLT) phase has been studied on the square-octagon lattice,  where three quantum critical points (QCPs) separate four phases in the model \cite{1987Rigorous}. By using a large-scale quantum Monte Carlo (QMC), it was shown that all three bulk QCPs belong to the same $O(3)$ universality class \cite{zhang2017unconventional}. However, the surface critical behaviors are different at each QCP due to the nontrivial surfaces. From the perspective of spectrum, the gapless mode plays an important role on the surfaces. So far, few works have focused on the spectrum. It is very helpful to study the dynamical properties of edge spins.
\begin{figure*}[t]
\centering
\includegraphics[width=0.96\textwidth]{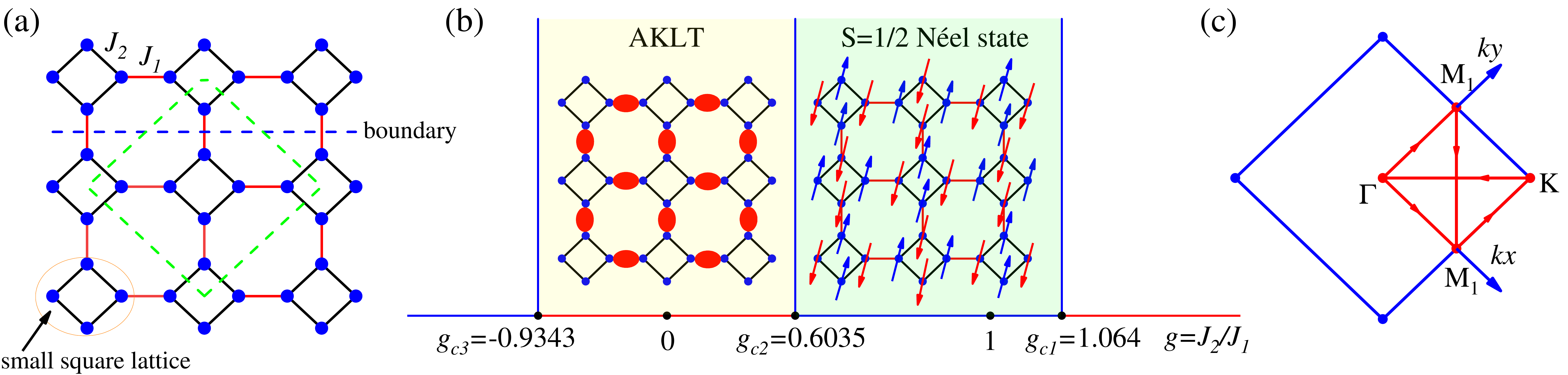}
\caption{(a) The square-octagon lattice. Within each unit cell (UC) there are four sites. The inter-UC Heisenberg coupling $J_{1}$ is antiferromagnetic and is set to be unity. The intra-UC Heisenberg coupling $J_{2}$ can be either ferromagnetic or antiferromagnetic.  The orange solid line denotes the unit cell containing four spins locating at a small square lattice. The green dotted line represents the doubled unit cell containing eight spins, and the blue dotted line cuts out the boundary of the lattice. (b) The quantum phase diagram obtained by tuning $g$ ($g_{c1}$: PVBC-N\'eel state QCP; $g_{c2}$: AKLT-$S=\frac 1 2$ N\'eel state QCP; $g_{c3}$: AKLT-$S=2$ N\'eel state QCP) \cite{zhang2017unconventional}. Two phases with three QCPs are only shown discussed here. The ellipses in the AKLT phase denote the spin singlets formed on the bonds. (c) The Brillouin zone of the square-octagon lattice and the high-symmetry points $\Gamma,M_{1},M_{2}$, and $K$.}
\label{fig:lattice}
\end{figure*}

Although the ground state of the model has been well understood, the dynamical properties of the SPT phase have rarely been studied. During the studies of spin excitations of the square lattice antiferromagnetic (AFM) Heisenberg model, a question arises for its high-energy spectrum anomaly at $q=(\pi,0)$ \cite{coldea2001spin,headings2010anomalous}. In this model, there is an obvious shift and large continuum at $q=(\pi,0)$ which can not be explained well by the linear spin wave theory (sw) \cite{singh1995spin,sandvik2001high}. Constructing the spinon picture \cite{shao2017nearly} can fit well with the numerical and experiment results. Powalski $et$ $al$. \cite{powalski2018mutually} proposed the strong attractive interaction between the spin wave modes to explain this anomaly phenomenon. These explanations demonstrate that strong correlation effects may exist in the high energy excitations of the N\'eel phase. Since the square-octagon lattice is very similar to the square lattice, it is necessary to check whether the anomaly can appear in its spectrum. In addition, it is not clear how the unconventional surface critical behavior influences the full spectra. In this paper, we focus on the dynamical properties of  the $S=1/2$ Heisenberg model on a square-octagon lattice. In a pure AKLT phase, it is well known that the edge spins are almost free. But what interests us is the dynamic behavior of bulk and edge spins when interaction and entanglement become stronger. In terms of numerical calculations, stochastic analytic continuation (SAC)\cite{beach2004identifying,sandvik2016constrained} with the QMC method \cite{sandvik1998stochastic,sandvik2010computational,sandvik2019stochastic,yan2019sweeping,yan2020improved} can reveal the spectrum information of spin models well. This QMC-SAC approach has gained the dynamical information of quantum magnets successfully from the square lattice checkerboard Heisenberg model \cite{xu2019spin} to quantum spin liquid \cite{sun2018dynamical,zhou2020string,zhou2020quantum,yan2021topological,zhou2021emergent}. Mainly, via this method assisted with linear spin wave approximation, the many-body perturbation method and symmetry analysis, we study the Heisenberg model on the square-octagon lattice to reveal its interesting dynamic properties.

\section{Model and methods}
The AFM Heisenberg model on the square-octagon lattice [Fig. \ref{fig:lattice} (a)] can be written as
\begin{equation}
\label{eq1}
H=J_{1}\sum_{\langle ij\rangle}\mathbf{S}_{i}\cdot \mathbf{S}_{j}+J_{2}\sum_{\langle ij\rangle'}\mathbf{S}_{i}\cdot \mathbf{S}_{j}
\end{equation}
where $\langle ij\rangle$ and $\langle ij\rangle'$ indicate the corresponding inter-UC bonds. Here, this unit cell only contains four spins on the small square as Fig. \ref{fig:lattice} (a) shows. The inter-UC coupling $J_{1}$ is antiferromagnetic and takes a fixed value of 1. We define the tuning parameter $g=J_{2}/J_{1}$ which can drive the model into four different phases~\cite{zhang2017unconventional}. Three QCPs denoted by $g_{ci}$, $i=1-3$ [Fig. \ref{fig:lattice} (b)] have been confirmed to be the $O(3)$ universal class analytically and numerically . As Fig. \ref{fig:lattice} (b) shows, the model is in the AKLT phase when $g_{c3}< g< g_{c2}$ \cite{1987Rigorous} and the $S=1/2$ N\'eel phase appears if $g_{c2}< g< g_{c1}$. In this paper, we mainly focus on the dynamical properties of the AKLT phase and the N\'eel phase separated by $g_{c2}$ which has different surface behaviors \cite{weber2018nonordinary,zhang2017unconventional}.

Via QMC simulations, we calculate the imaginary-time correlation functions under different parameter $g$s from the AKLT phase to the N\'eel phase in the model. The lattice size $L$ is chosen to be $24$ with the periodic boundary condition (PBC) and $32$ with the open boundary condition (OBC). Inverse temperature $\beta=2L$ is used for the QMC calculation. With the fully periodic boundary condition, two kinds of bulk spin-spin imaginary-time correlation functions are measured in the QMC simulations. One is the bulk spin-$\frac {1} {2}$ spin-$\frac {1} {2}$ correlation function $G_{b(1/2)}(\mathbf{q},\tau)=\frac{1}{L^2}\sum_{i,j}e^{-i\mathbf{q}\cdot(\mathbf{r}_{i}-\mathbf{r}_{j})}\langle s^{z}_{i}(\tau)s^{z}_{j}(0)\rangle$, another is the bulk effective spin-2 spin-2 correlation function $G_{b2}(\mathbf{q},\tau)=\frac{1}{L^2}\sum_{i,j}e^{-i\mathbf{q}\cdot(\mathbf{r}_{i}-\mathbf{r}_{j})}\langle S^{z}_{i}(\tau)S^{z}_{j}(0)\rangle$. Here, $s^{z}_{i}$ is a real spin-$\frac 1 2$ located at the lattice point. $S^{z}_{i}$ is a spin-2 made up of four spin-$\frac 1 2$s satisfying $S^{z}_{i}=\sum^{4}_{n=1}s^{z}_{in}$ on a small square lattice $i$ as shown in the left bottom of Fig.\ref{fig:lattice} (a). Although it is not an exact spin 2, it can be considered physically qualitative in this way.

Furthermore, we measure two kinds of spin-spin correlation functions on the boundary as the blue dashed line of Fig. \ref{fig:lattice} (a) shown, including spin-$\frac{1}{2}$ spin-$\frac{1}{2}$ correlation $G_{e(1/2)}(\mathbf{q},\tau)$ and spin-2 spin-2 correlation $G_{e2}(\mathbf{q},\tau)$ to investigate dynamical information of spin on the edge. They can be expressed as
$G_{e(1/2)}(\mathbf{q},\tau)=\frac{1}{L}\sum_{i,j}e^{-i\mathbf{q}_{x}\cdot(\mathbf{x}_{i}-\mathbf{x}_{j})}\langle s^{z}_{i}(\tau)s^{z}_{j}(0)\rangle$ and $ G_{e2}(\mathbf{q},\tau)=\frac{1}{L}\sum_{i,j}e^{-i\mathbf{q}_{x}\cdot(\mathbf{x}_{i}-\mathbf{x}_{j})}\langle S^{z}_{i}(\tau)S^{z}_{j}(0)\rangle$.
 After obtaining the imaginary-time correlation functions, the SAC algorithm will be employed to find the suitable spectrum functions $S_{b(1/2)}(q,\omega)$, $S_{b2}(q,\omega)$, $S_{e(1/2)}(q,\omega)$ and $S_{e2}(q,\omega)$ to fit the above imaginary-time correlation functions within the stochastic errors via the Monte Carlo sampling methods \cite{sandvik2016constrained,shao2017nearly}.\\
 \begin{figure*}[htp!]
 \centering
 \includegraphics[width=0.85\textwidth]{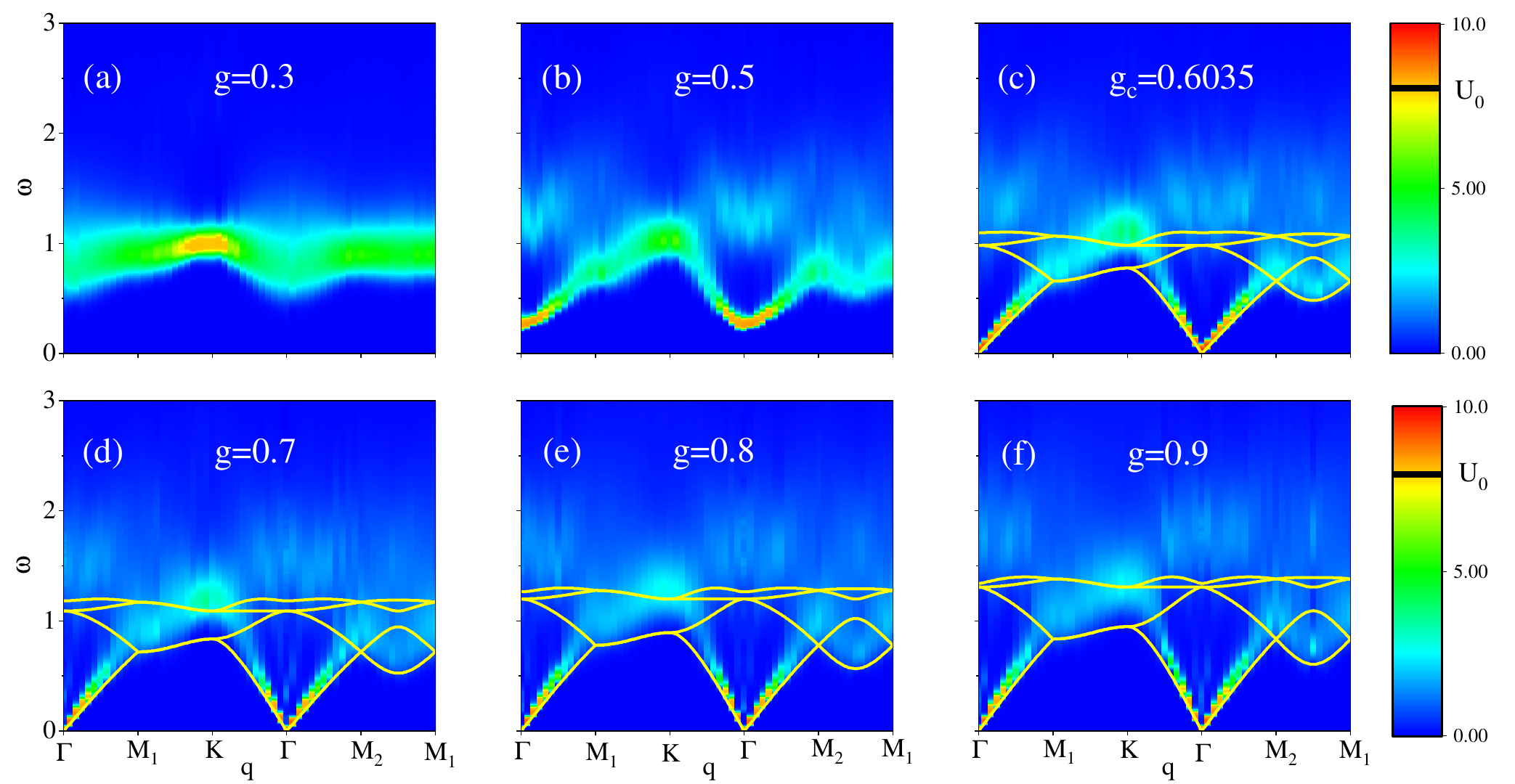}
 \caption{Bulk spin-$\frac{1}{2}$ spectra $S_{b(1/2)}(q,\omega)$ with PBC obtained from QMC-SAC in different $g$s, where $g$ is (a) 0.3, (b) 0.5, (c) $g_{c}$(0.6035), (d) 0.7, (e) 0.8 and (f) 0.9. The yellow lines in the N\'eel phase (also including the QCP) are the results of linear spin-wave theory. In the color bar, the black line denotes the boundary between the linear and the logarithmic color mappings of the spectrum function. Below the boundary, the spectrum weight $U=S(q.\omega)$, whereas above the boundary $U=U_{0}+\log_{10}[S(q,\omega)/U_{0}]$ and $U_{0}=8$.}
 \label{Fig.smallspin}
 \end{figure*}

 \begin{figure*}[htp!]
 \centering
 \includegraphics[width=0.85\textwidth]{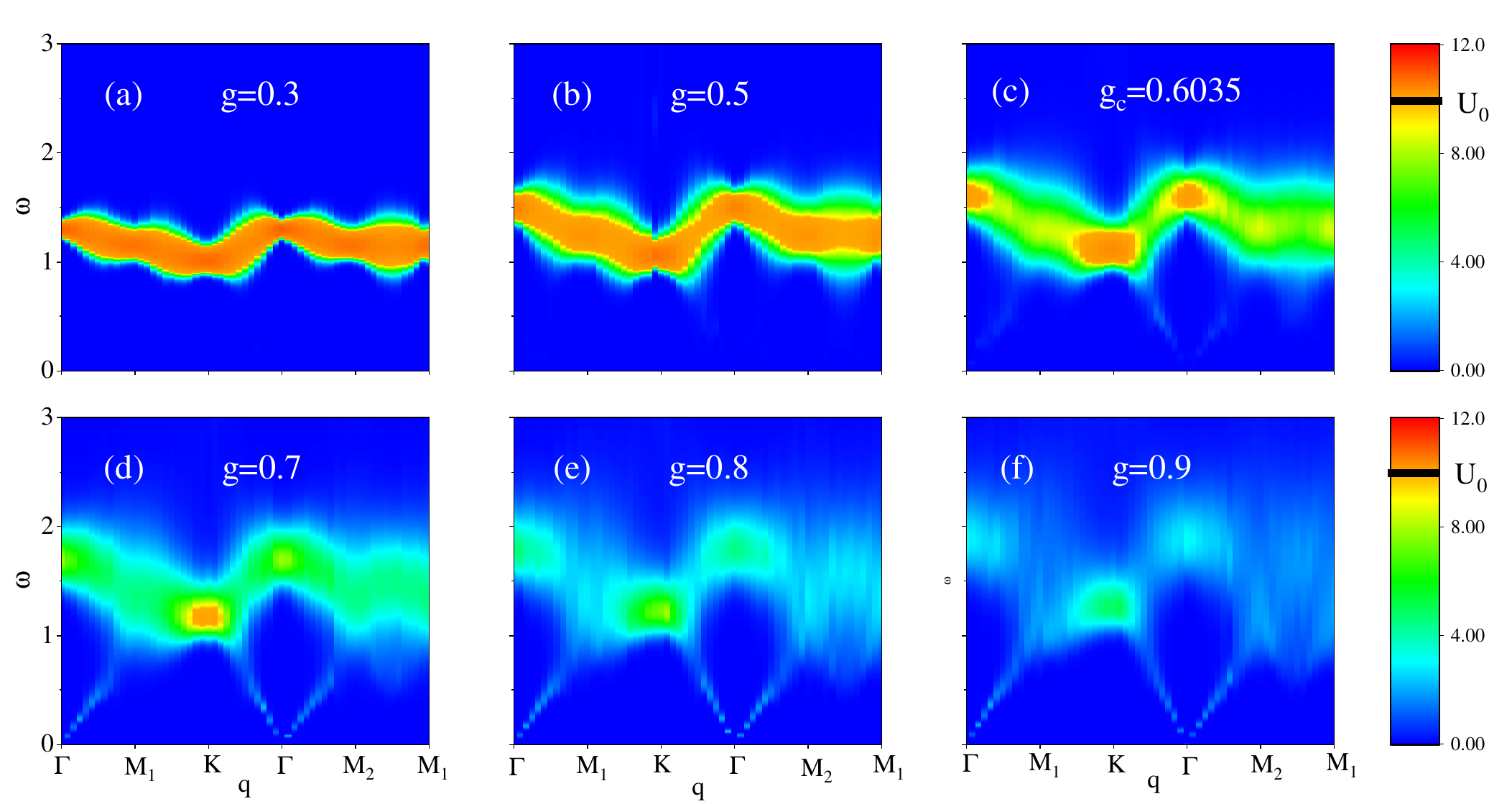}
 \caption{Bulk spin-2 spectra $S_{b2}(q,\omega)$ with the PBC obtained from QMC-SAC in different $g$s, where $g$ is (a) 0.3, (b) 0.5, (c) $g_{c}$(0.6035), (d) 0.7, (e) 0.8 and (f) 0.9. The boundary between the linear and the logarithmic color mappings is $U_{0}=10$.}
 \label{Fig.bigspin}
 \end{figure*}

 \section{NUMERICAL RESULTS}
 \subsection{Bulk spin spectra}

 First, we consider the bulk spin spectra with the PBC. In order to compare the bulk spectra with linear spin-wave results, eight spins in each unit cell are chosen as green dashed lines that Fig. \ref{fig:lattice} (a) shows. The results of $S(q,\omega)$ are shown in Fig. \ref{Fig.smallspin} and \ref{Fig.bigspin} , and the path we choose is $\Gamma(0,0) \rightarrow M_{1}(\pi,0) \rightarrow K(\pi,\pi) \rightarrow \Gamma(0,0) \rightarrow M_{2}(0,\pi) \rightarrow X_{1}(\pi,0)$ as Fig. \ref{fig:lattice} (c) shows. At $g=0.3$, the ground state is the AKLT  phase, and its spin excitation is gapped at the $\Gamma$ point. Obviously, a flat band arises in the AKLT phase as Fig. \ref{Fig.smallspin} (a) shows ($g=0.3$); it means the spins couple into singlets and  become localized. When $g$ increases, the spin gap at $\Gamma$ gradually closes which supports a continuous phase transition. The spin excitation keeps gapless in the whole N\'eel phase due to the Goldstone theorem \cite{goldstone1962broken,sachdev2011quantum}. Indeed, we observe the gapless Goldstone mode at the $\Gamma$ point as Figs. \ref{Fig.smallspin} (c)-(f) show.

 For the spectra of the spin-2 case, the distribution of spectra weight is different from the spin-$\frac{1}{2}$ case. In the N\'eel phase, the upper band has main weight, and the Goldstone mode looks much weaker than the spin-$\frac{1}{2}$ case. The low-energy excitation is the Goldstone mode caused via the spin-wave mechanism, that is similar for both small and large spin excitations. The excitation of large spin $S^{z}_{i}=\sum^{4}_{n=1}s^{z}_{in}$ contains not only the information of the same index $n$ of spin $\frac{1}{2}$, but also the information of different index $n$ of spin $\frac{1}{2}$. Whereas the spectrum of small spin $\frac{1}{2}$ only reveals the dynamical property of the same index of spin, so its higher band is weaker than the spin-2 case.

 To better understand the bulk spin spectra in the N\'eel phase, we use linear spin-wave theory to calculate the low-energy branches of the model as shown in Fig. \ref{Fig.smallspin}. Assuming the ground state is the N\'eel order, a Holstein-Primakoff transformation transforms the spin operators at the linear-wave level \cite{toth2015linear}, which are expressed in terms of boson creation and annihilation operators
 $S^{z}_{i}=S-a^{+}_{i}a_{i}$, $S^{+}_{i}\approx \sqrt{2S}a_{i}$, $ S^{-}_{i}\approx \sqrt{2S}a^{+}_{i}$, and
 $S^{z}_{j}=b^{+}_{j}b_{j}-S$, $S^{+}_{j}\approx \sqrt{2S}b^{+}_{j}$,  $S^{-}_{j}\approx \sqrt{2S}b_{j}$,
 where $a^{+}_{i}$ and $a_{i}$ are for up spin and $b^{+}_{j}$ and $b_{j}$ are for down spin. The linear spin-wave Hamiltonian $H_{sw}$ can be obtained by using the Holstein-Primakoff transformation. And then $H_{sw}$ will be transformed into the momentum space by Fourier transformation and numerically diagonalized to get the spin-wave dispersions. There are four spin-wave branches in the N\'eel phase, each of which is actually twofold degenerate due to the double UCs. When $g>g_{c}$, the spin-wave results match well with the numerical results in the low-energy part. Even though $g=g_{c}$, the spin-wave results also capture the low-energy branches. Besides, the high-energy excitations of spectra obtained from QMC-SAC can be partially explained by two high-energy spin-wave branches qualitatively. It shows that the interaction between these two spin-wave modes can not be ignored which makes the high-energy bands wide and undulating. In addition, the low-energy part of spin-wave dispersions of the spin-$\frac{1}{2}$ case(Fig.\ref{Fig.smallspin}) can fit well with the low-energy part of the  spin-2 case (Fig.\ref{Fig.bigspin}) in the numerical results, denoting that the dynamical properties of the spin-$\frac{1}{2}$ case can also describe the spin-2 case. At the $K$ point, spin-wave dispersion has a shift which is similar to the QMC-SAC results, but the deviations are different from each other. Two magnon dispersions overlap from $M_{1}$ to $K$ so that their strong interaction may lead to a dispersion shift.
 Unlike the spin-$\frac{1}{2}$ Heisenberg model on the square lattice \cite{shao2017nearly}, we have not seen the obvious spinon continuum in the N\'eel phase on the square-octagon lattice. In this model, the phase-transition point between the N\'eel phase and the AKLT phase belongs to the $O(3)$ universal class instead of being a deconfined quantum critical point, so it may make spinons confined with continuum disappearing. Around the $\Gamma$ point, there is a higher and weaker gapped mode beyond the Goldstone mode as shown in Fig. \ref{Fig.smallspin}. It is obviously different from the two-magnon excitation according to the scale and value. This is probably the famous Higgs amplitude mode which happens on the module of the order parameter~\cite{zhou2021amplitude}. The similar phenomenon was also found in the N\'eel phase of similar Heisenberg systems with phase transition between the valence bond phase and the N\'eel phase in experiment~\cite{2017Higgs}.

 \subsection{Edge spin spectra}

\begin{figure*}[htp]
 \centering
 \includegraphics[width=0.85\textwidth]{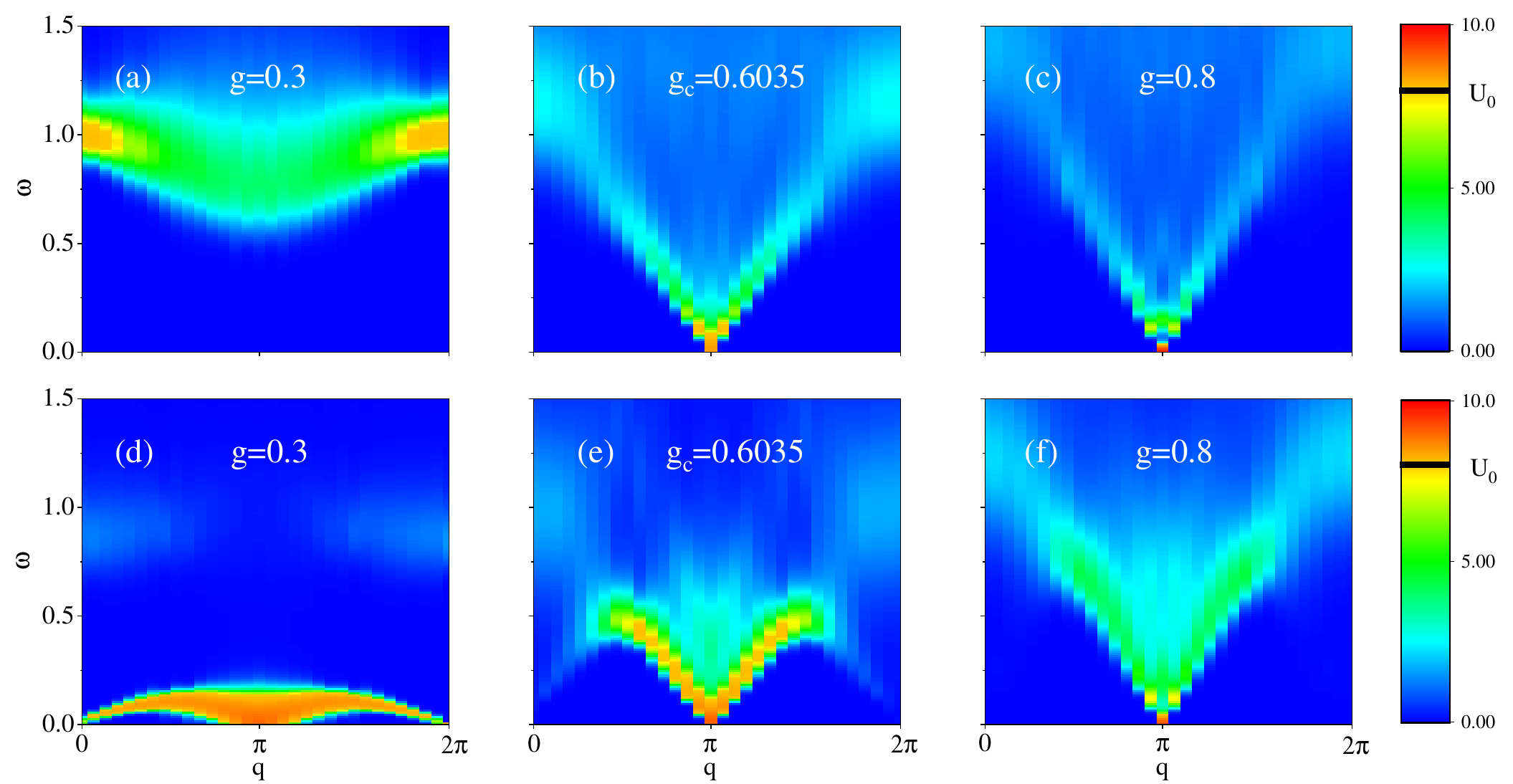}
 \caption{Edge spin-$\frac{1}{2}$ spectra obtained from QMC-SAC in different $g$s, where $g$ is (a) and (d) 0.3, (b) and (e) $g_{c}$(0.6035), and (c) and (f) 0.8. (a)-(c) are spectra $S_{e(1/2)}(q,\omega)$ with the PBC whereas (d)-(f) are spectra $S_{e(1/2)}(q,\omega)$ with the OBC. The boundary between the linear and the logarithmic color mappings is $U_{0}=8$.}
 \label{Fig.edgesmall}
 \end{figure*}

 Under the PBC, the properties of phases and phase transitions can be picked up from dynamical properties of spin on the boundary (although with the PBC, this lattice has no real edge, we compare the same one-dimensional (1D) spin chain in this system with the edge under the OBC). By using the 1D Fourier transform of the spin $\frac{1}{2}$ or spin 2 along the boundary, we get the spectra of the edge spin. For $g=0.3$, the spin excitation is gapped in the spin-$\frac{1}{2}$ spectra. If $g \rightarrow 0$, the gap becomes larger, and it becomes a flatter band at $\omega=1$ with $g=0$. When $g$ keeps increasing, the gap gradually closes at $q=\pi$ [Fig. \ref{Fig.edgesmall} (c)], which is similar to what we observe in the bulk spin spectra. This is the characteristic of the second order phase transition that can also be detected by the spectra of edge spins. For $g=g_{c}$ and $g=0.8$, the spin excitation becomes linear and gapless at $q=\pi$ as Figs. \ref{Fig.edgesmall} (b) and \ref{Fig.edgesmall} (c) show. According to the Goldstone theorem, a gapless mode with linear dispersion can be observed at $q=\pi$ in the AFM phase which is consistent with our calculation. The two magnon modes attribute to the linear spin excitations at $q=\pi$ \cite{ma2018dynamical} {\color{red}}. The change in spectra agrees well with the theoretical predictions.

 Edge spin-2 spectra show that a nearly flat band appears at $g=0.3$ which is similar to spin-$\frac{1}{2}$ excitation. But weak gapless modes appear at $q=0$ and $q=2\pi$ together with the obvious high-energy excitation in the N\'eel phase (for spin 2, the N\'eel phase actually becomes a ferromagnetic phase).
 \begin{figure*}[htp]
 \centering
 \includegraphics[width=0.85\textwidth]{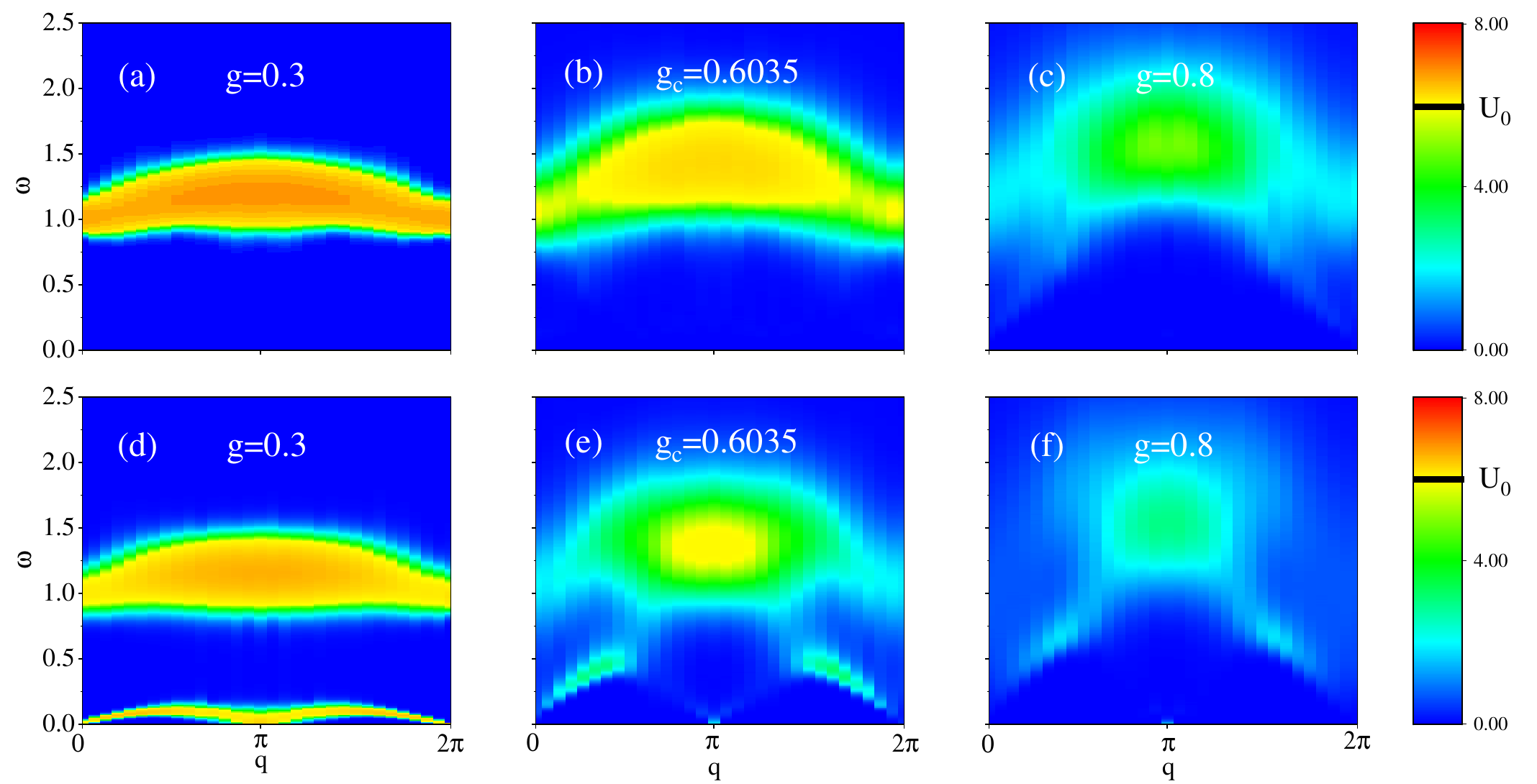}
 \caption{Edge spin-2 spectra obtained from QMC-SAC in different $g$s, where $g$ is (a) and (d) 0.3, (b) and (e) $g_{c}$(0.6035), and (c) and (f) 0.8. (a)-(c) are spectra $S_{e2}(q,\omega)$ with the PBC whereas (d)-(f) are spectra $S_{e2}(q,\omega)$ with the OBC. The boundary between the linear and  the logarithmic color mappings is $U_{0}=6$.}
 \label{Fig.edgebig}
 \end{figure*}
Now, let us focus on the real edge spin spectra with the open boundary condition. At $g=0.8$, where the model is in the deep N\'eel phase, the spin-$\frac{1}{2}$ excitation is almost similar to the case with the PBC, which keeps linear dispersion and has no gap at $q=\pi$ [Fig.\ref{Fig.edgesmall}(f)]. However, when $g$ keeps decreasing, we do not observe that the gap opens at the $\pi$ point. Instead, the gap closes not only at the $\pi$ point but also at the $\Gamma$ point, which is different from the free spin chain and edge spins with the PBC. Clearly, the spin excitation becomes an arched continuum in the lower-energy part [Fig.\ref{Fig.edgesmall} (d)]. This suggests these edge spins enter an effective Luttinger liquid phase, causing the gap closing at the $\pi$ and $\Gamma$ points. Due to the $SU(2)$ symmetry of the edge chain, we suppose that it is an effective $S=1/2$ Heisenberg chain on edges instead of a XX spin chain. From the Fig. \ref{Fig.edgesmall} (e), we find that there are no well-defined magnon modes at $q=\pi$. The arched two-spinon continuum gradually emerges and separates from the weak high-energy excitations. Around $q=\pi/2$ and $q=3\pi/2$, the spin-$\frac{1}{2}$ excitation splits into two parts at $\omega \sim 0.7$ with $g=g_{c}$, which is caused by edge spins with the OBC. The lower-energy branch can be described by the effective $S=1/2$ Heisenberg chain, and the boundary of spinon continuum depends on the amplitude of $J_{2}$. If $J_{2}$ approaches zero, this low-energy branch becomes nearly flat at $\omega=0$.

In the spectra of the spin-2 case, the low-energy gapless mode as shown in Fig. \ref{Fig.edgebig} (e) is similar to the spin-$\frac{1}{2}$ case which emerges an effective Luttinger liquid phase in the low-energy part at $g=0.3$. And as Figs. \ref{Fig.edgebig} (e) and \ref{Fig.edgebig} (f) show, the low-energy part of the spectrum weight is weak at $q=\pi$ which still carries features of the spin-$\frac 1 2$ N\'eel phase. With the OBC, the spin-$\frac 1 2$ on the edge is more free and not equivalent to other spin $\frac 1 2$s, so the weak gapless mode of the spin $\frac 1 2$ appears at $q=\pi$ in the spectrum. The appearance of the high energy branch also deserves theoretical analysis. According to Fig. \ref{Fig.edgesmall} (d), there is another high-energy branch at $\omega=1$ above the two-spinon continuum.  It can be related to the triplet excitation since it locates at $\omega=1$, and we will argue this analytically during the following discussion.

 \section{DISCUSSION}
 In the SPT phase, the Lieb-Schultz-Mattis (LSM) theorem \cite{lieb1961two,oshikawa2000commensurability,hastings2004lieb} applies here and defines the feature of the spectra.  When the LSM theorem generalizes to the SPT phase, it shows that the ground state of the $(d+1)$-dimension bulk state of the SPT phase must be gapped and nondegenerate, whereas the $d$-boundary state must be either gapless or degenerate \cite{jian2018lieb}. For $g=0.3$, the symmetry is preserved on the boundary, so the spectrum is gapless according to the LSM theorem [Figs. \ref{Fig.edgesmall}(f) and \ref{Fig.edgebig}(f)].
 \begin{figure}[h]
 \centering
 \includegraphics[width=0.45\textwidth]{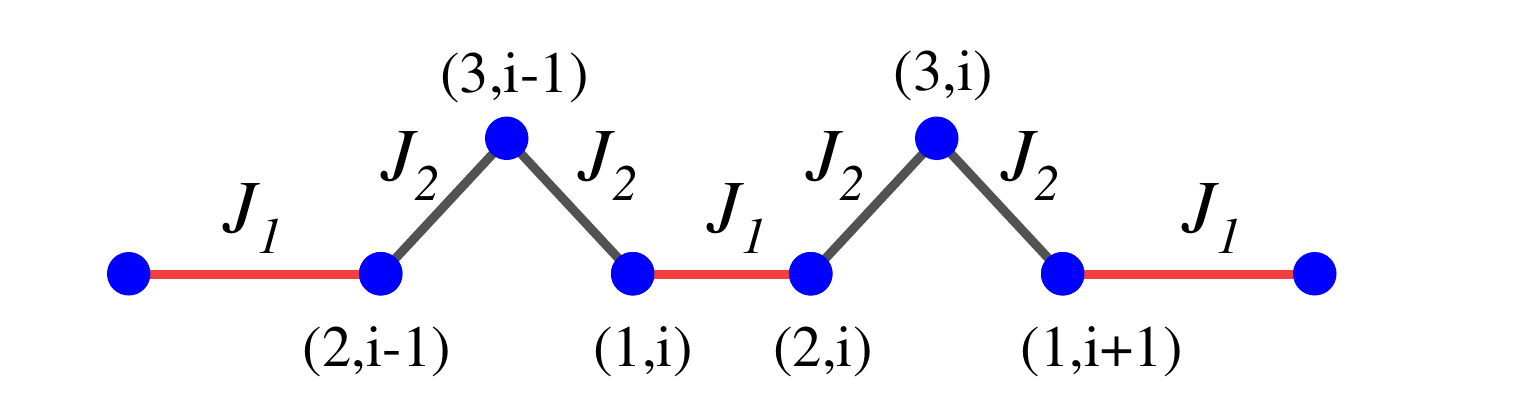}
 \caption{ A spin-$\frac{1}{2}$ trimerized Hesienberg chain cut from the boundary of the decorated square lattice. }
 \label{Fig.Heff}
 \end{figure}

  The spins on the boundary can be approximately treated as a $S=1/2$ spin chain in order to understand its Luttinger-liquid-like spectrum on the boundary better. Assuming the interaction between the edge spin $\frac{1}{2}$ and the bulk spin $\frac{1}{2}$ is weak enough in the deep AKLT phase, the edge spin $\frac{1}{2}$s form a trimerized Heisenberg chain as Fig. \ref{Fig.Heff} shows. In order to explain the lower excitation of edge spins better, we will use the many-body perturbation method \cite{cheng2022fractional,verkholyak2021modified} to get the effective Hamiltonian (for details, see Ref \cite{verkholyak2021modified}). The trimerized Heisenberg chain is given by the following Hamiltonian:
 \begin{equation}
 \label{eq2}
 H=\sum^{N}_{i}[J_{1}\mathbf{S}_{1,i}\cdot \mathbf{S}_{2,i}+J_{2}(\mathbf{S}_{2,i}\cdot \mathbf{S}_{3,i}+\mathbf{S}_{3,i}\cdot \mathbf{S}_{1,i+1})]
 \end{equation}
 where $J_{1}$ and $J_{2}$ are the same as the square-octagon lattice. The effective Hamiltonian would describe the effective interaction between $\mathbf{S}_{3,i}$ and $\mathbf{S}_{3,i+1}$. Supposing $J_{2}$ $\ll$ $J_{1}$, the perturbation theory starts from the noninteracting dimer $\mathbf{S}_{1,i}$-$\mathbf{S}_{2,i}$ and monomer $\mathbf{S}_{3,i}$.
 And the initial Hamiltonian can be divided into two parts $H=H_{IH}+V$ . By means of the second-order perturbation expansion, the initial Hamiltonian $H=H_{IH}+V$ can be projected into a certain subspace.  So the final effective Hamiltonian \cite{verkholyak2021modified} is shown in Eq. (\ref{eq3}),
 \begin{align}
 \begin{split}
 \label{eq3}
 H_{eff}&=J_{eff}\sum^{N}_{i}\mathbf{S}_{3,i}\cdot \mathbf{S}_{3,i+1}  \\
 J_{eff}&=\frac{J^{2}_{2}}{2J_{1}-J_{2}}
 \end{split}
 \end{align}

\begin{figure}[htp]
 \centering
 \includegraphics[width=0.44\textwidth]{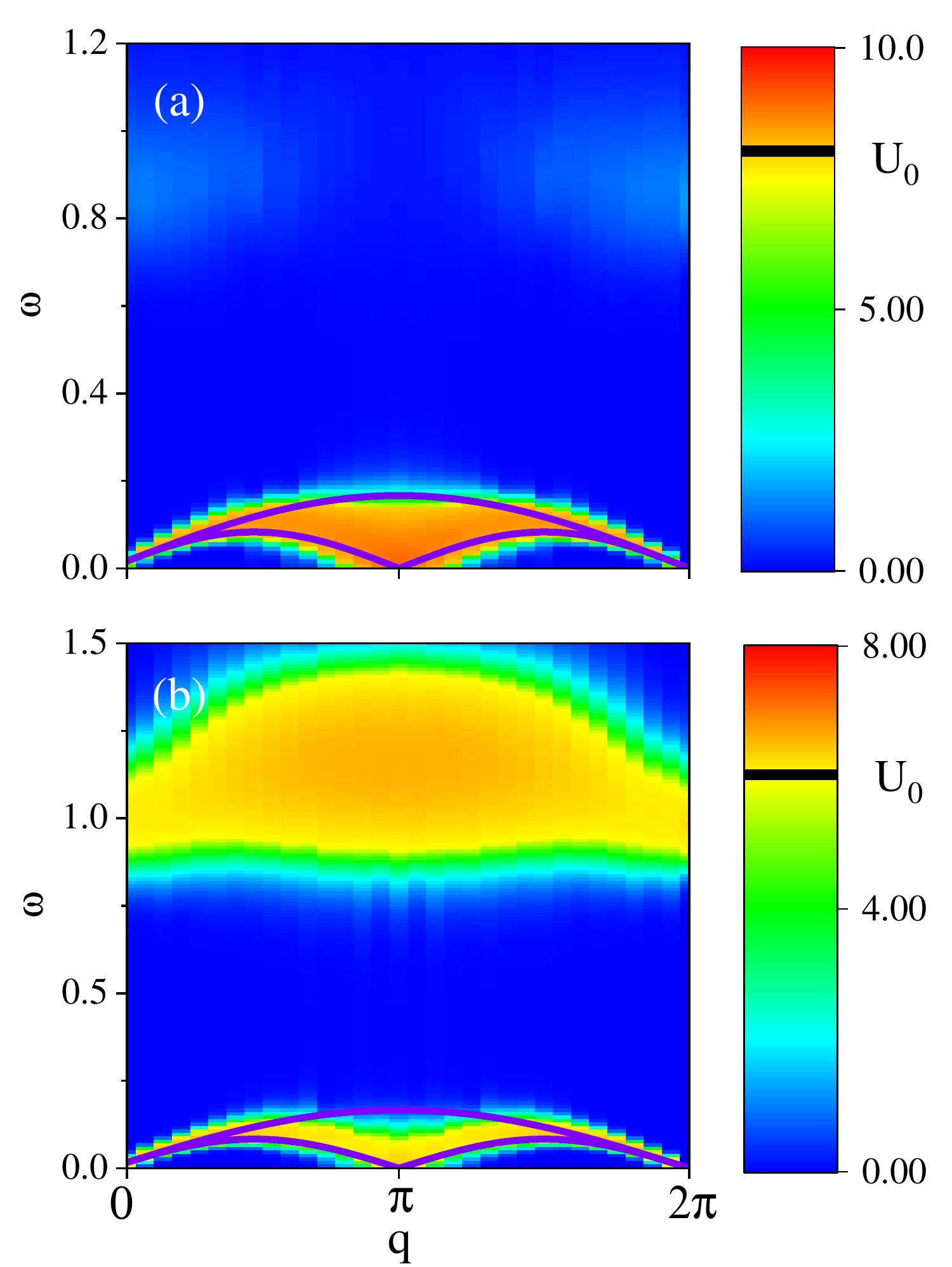}
 \caption{Effective Hamiltonian results compared with spectra obtained from QMC-SAC. (a) and (b) are the full spectra of spin $\frac{1}{2}$ and spin 2 on the boundary with $g$=0.3. The purple lines in (a) and (b) are the results of the many-body perturbation approximation based on spin $\frac{1}{2}$. The boundary between the linear and the logarithmic color mappings is (a) $U_{0}=8$ and (b) $U_{0}=6$. }
 \label{Fig.per}
 \end{figure}

 According to the Bethe ansatz solution of the $S=1/2$ Heisenberg chain \cite{2006The}, the upper and lower boundaries of the spectrum for the effective Heisenberg Hamiltonian are shown as purple lines in Figs. \ref{Fig.per} (a) and \ref{Fig.per} (b). They satisfy the following functions $\omega_{lower}=\pi J_{eff}$ $\vert$sin($q$)$\vert$/2 and  $\omega_{upper}=\pi J_{eff}$$\vert$sin($q$/2)$\vert$.  Although edge spins interact weakly with bulk spins, the boundaries of the spectrum for the effective Hamiltonian could meet the results of QMC-SAC qualitatively. Although the effective Hamiltonian is derived in the spin-$\frac{1}{2}$ Hilbert space, it can explain successfully the low-energy excitation of spin 2 as Fig. \ref{Fig.per} (b) shows. In addition, the intradimer ($J_{1}$ bond) excitation will appear in the high-energy part because triplon excitation from the $J_{1}$ bonds would move along the chain with nonzero $J_{2}$. If we calculate the spectra of the effective spin $\frac{1}{2}$ in the spin-$\frac{1}{2}$ trimer chain, there will be triplon excitation at $\omega \sim 1$. Therefore, the weak triplet excitation can also be found in the edge spin-$\frac{1}{2}$ spectra at $\omega \sim 1$ as Fig. \ref{Fig.per} (a) shows, which is beyond the description of the low energy effective Hamiltonian. Because these dimers of $J_{1}$  attribute to the spin-2 spin-2 correlation and the edge spin 2s are strongly coupled to the bulk spin 2s, the triplon dispersion at $\omega=1$ is quite strong and obvious in Fig. \ref{Fig.per} (b). And this high-energy band can not be explained by the low-energy effective Hamiltonian.

\section{CONCLUSION}
To summarize, we utilize QMC-SAC assisted with linear spin-wave theory, the many-body perturbation method, and symmetry analysis to study both bulk and edge dynamical properties of the $J_1$-$J_2$ AFM Heisenberg model on the square-octagon lattice. From the flat band of the AKLT phase to the linear mode of the N\'eel phase, the gap at $\Gamma$ gradually closes in the bulk spin spectra. Comparing with spin-wave theory, we find that the low-energy branch can be captured by the spin-wave results but the strong correlation effect is observed for the high-energy branches. Besides, we also see a possible Higgs amplitude mode in the N\'eel phase which is similar to experiment \cite{2017Higgs}. On the edges, an emergent Luttinger liquid phase with a nearly flat band is observed in the AKLT phase. Furthermore, we obtain an effective 1D AFM Heisenberg Hamiltonian from many-body perturbation method to explain this low-energy excitation qualitatively. The rich physics of this model helps us to better understand the dynamical behavior of the SPT phase and magnetic orders.

\section{Acknowledgements}
We wish to thank L. Zhang, H.-Q. Wu, C. Liu, S.-N. Ning, J. Feng, Z. Xiong, and C. Zhou for helpful discussions. Z.L., J.L., J.L., and D.-X.Y. are supported by NKRDPC-2017YFA0206203, NKRDPC-2018YFA0306001, NSFC-11974432, NSFC-92165204, NSFC-11804401, NSFC-11832019, GBABRF-2019A1515011337, Leading Talent Program of Guangdong Special Projects, and Shenzhen Institute for Quantum Science and Engineering (Grant No. SIQSE202102). R.-Z.H. is supported by China Postdoctoral Science Foundation (Grant No.2020T130643), Fundamental Research Funds for the Central Universities, and National Natural Science Foundation of China (Grant No. 12047554).

%\bibliography{AKLT}

\end{document}